\documentclass{article}
\usepackage[utf8]{inputenc}
\usepackage{graphicx}
\usepackage{authblk}
\usepackage{url}
\usepackage{listings}
\title{The Constitutions of Web3}

\author[1,2]{Joshua Tan}
\author[1]{Max Langenkamp}
\author[3]{Anna Weichselbraun}
\author[4]{Ann Brody}
\author[1]{Lucia Korpas}
\affil[1]{Metagov}
\affil[2]{University of Oxford}
\affil[3]{University of Vienna}
\affil[4]{McGill University}

\begin{document}
\maketitle

\begin{abstract}
The governance of online communities has been a critical issue since the first USENET groups, and a number of serious constitutions---declarations of goals, values, and rights---have emerged since the mid-1990s. More recently, decentralized autonomous organizations (DAOs) have begun to publish their own constitutions, manifestos, and other governance documents. There are two unique aspects to these documents: they (1) often govern significantly more resources than previously-observed online communities, and (2) are used in conjunction with smart contracts that can secure certain community rights and processes through code. In this article, we analyze 25 DAO constitutions, observe a number of common patterns, and provide a template and a set of recommendations to support the crafting and dissemination of future DAO constitutions. We conclude with a report on how our template and recommendations were then used within the actual constitutional drafting process of a major blockchain.
\end{abstract}

\section{Introduction}
A decentralized autonomous organization, or DAO, is an online community organized through one or more smart contracts. Some of these smart contracts are quite straightforward, with only a treasury. Others are more sophisticated, with governance processes such as quadratic voting or conditional timelocks.

Smart contracts mark a departure from traditional governance documents. Whereas traditional documents such as textual constitutions or policy statements can declare rights and rules to be enforced through some other mechanism, a smart contract can directly provision those rights and rules. For example, if xDAO creates a smart contract that requires at least five keys to approve a proposal, all proposals to xDAO with less than five people will be automatically rejected.

Although smart contracts are promising and increasingly important, they alone cannot govern communities; traditional constitutions and declarations of rights are still crucial for good governance. First, rights within smart contracts tend to only be relevant for a small set of transactions, each of which requires highly-legible data. Currently, it is neither feasible nor (arguably) desirable to encode all rights, rules, and processes within smart contracts. Second, even when rights can be incorporated within smart contracts, doing so renders them illegible to those participants who cannot read the code within the smart contract. In other words, rights in smart contracts also need to be articulated with accompanying texts. 

This essay aims \begin{enumerate}
    \item to understand and contextualize the usage of written constitutions and smart contracts within the emerging politics of DAOs and other digitally-constituted organizations, 
    \item to analyze these textual documents in their own right, and 
    \item to develop a stable set of principles and practices for future DAO constitutions.
\end{enumerate}

The paper proceeds in four parts. In Part I, we review the history of digital constitutionalism on the internet. In Part II, we describe the rights, values, and goals of constitutions of Web3 and analyze some of the stylistic and political patterns we found. In Part III, we present recommendations, along with a technical publication standard, for future DAO and Web3 constitutions. In Part IV, we identify future directions for this research, focusing on work to analyze existing governance smart contracts to understand the types of rights and affordances that have (and have not) been written into them.

\section{Part I: Digital constitutionalism and Web3}
The early internet hosted a number of notable constitutions and governance proposals. For example, Redeker, Gil, and Gasser\cite{redeker2018} collected and analyzed an extraordinary collection of constitutions and manifestos over the history of the internet, ranging from the early \textit{Rights and Responsibilities of Electronic Learners} in 1992 to 2014’s \textit{Magna Carta for Philippine Internet Freedom}. They describe digital constitutionalism as “a constellation of initiatives that have sought to articulate a set of political rights, governance norms, and limitations on the exercise of power on the Internet” \cite{redeker2018}.

Redeker et al.’s data set focuses on constitutions that assert rights and norms over the entire internet and/or key aspects of its infrastructure. Despite their emphasis on rights, most of these documents can be better understood as political declarations rather than as authoritative governance documents. By contrast, online communities throughout the history of the internet have enacted their own constitutions, which are more comparable to the governance documents of neighborhood associations or the charters of corporations. For example, early USENET groups (similar to many forum-based communities today) often posted a charter that described the intention and governance of that community \cite{newsdemon2022}, and open-source software communities such as Python have adopted similar governance procedures that articulate the rules and expectations for participating in these communities in comprehensive detail.

More recently, the development of blockchains and other cryptographic protocols have enabled the rise of a series of services and platforms collectively referred to as Web3. These technologies revolve around trustless mechanisms for social, economic, and political coordination. Many proponents of Web3 like to emphasize that typical forms of human governance (such as votes and constitutions) are not necessary or even desirable in Web3 communities. Other proponents of Web3 emphasize the importance of maintaining human-to-human mediation in addressing coordination problems. Despite the range of perspectives on how governance is best done, it is clear that group coordination requires shared goals and values. Constitutions can thus be understood as an interface for a community’s shared values.
	
Decentralized autonomous organizations, or DAOs, have become an increasingly prominent form of digital institution in Web3 \cite{zargham2021}. While the term is used differently in different communities, all DAOs embody certain institutional rules within code. 

Taking inspiration from Redeker et al.’s survey of internet constitutions, we study governance in Web3 through the lens of digital constitutionalism. What follows is a set of preliminary findings from an analysis of the governance documents from 19 different decentralized organizations.

\subsection{Data set and methods}
In gathering the raw data, we collected publicly-available constitutions and governance documents from a range of representative DAOs including large protocol DAOs, service DAOs, investment DAOs, and NFT DAOs, with an emphasis on covering a wide range of institutional types rather than the biggest or most well-known DAOs. Our data collection was limited by the fact that many major DAOs do not have obvious constitutions, though almost all describe some form of governance process beyond the smart contract. The raw data set is housed in a publicly available GitHub repository.

In cleaning and coding the raw dataset, we focused on capturing goals, values, and rights. Notably, we decided against trying to encode the rules and processes described in these documents not only because they were less salient in the documents we studied (see Part II) but because many of these young communities were still in the active process of determining or changing those rules and processes. We also incorporated several coding fields from the Comparative Constitutions Project \cite{elkins2010}, including `transprov`, `is\_Amend`, and `in\_Force`, allowing direct comparisons between DAO constitutions and national constitutions along a number of dimensions.  Other metadata include the title of the document, the date the constitution was created, aspects of its structure, and so on.

\begin{figure}[htbp]
\centering
\includegraphics[width=0.7\linewidth]{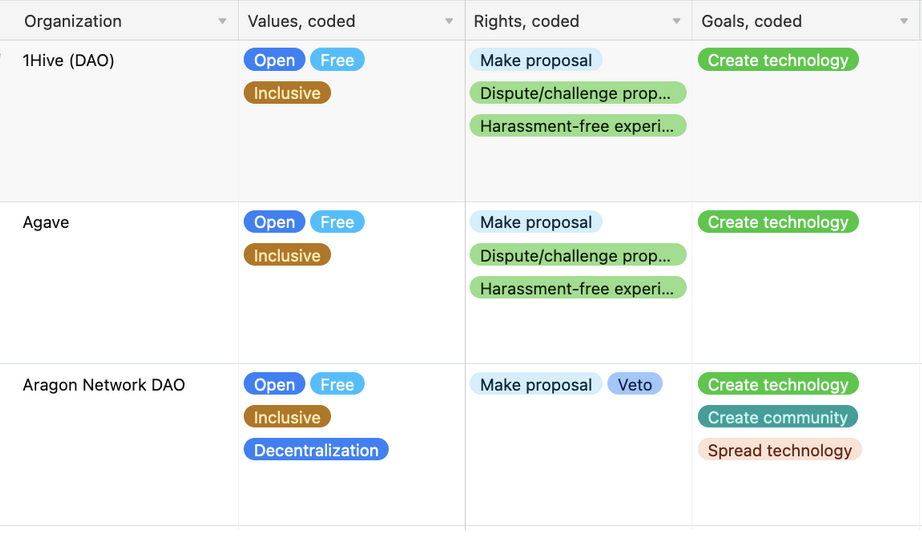}
\caption{Example of the coded information.}
\label{fig:mylabel}
\end{figure}

In coding for values, goals, and rights within a constitution, we first divided the document into sections following the top-level headers in the document, then matched different coded categories to each section (e.g. values are addressed in the “Our Values” section of the BrightDAO Charter, while rights are referenced in the “Our Mission”, “Our Vision”, and “Our Values” sections). In coding for these fields, we first collected the explicit words and descriptions used in the documents (e.g. “self-sovereign” or “sustainable decentralization”) and then mapped similar terms to a more restricted set of labels (e.g. “open”, “free”, “inclusive”, “professionalism”, and “decentralization”).

\section{Part II: Analyzing DAO constitutions}

Open-source communities have a history of debating the terms of their engagement and arguing about what it is they are doing, as well as producing governance documents from contributor guidelines to governance.md files—practices which in turn bring the community into being. These documents give us a window into how communities imagine themselves and how they grapple with questions of purpose and coordination.

Governance documents include manifestos, constitutions, codes of conduct, community covenants, charters, and other genres with overlapping but distinct functions. These documents all outline a set of common values and goals with varying degrees of explicitness. They share many features with earlier texts within online groups ranging from open-source communities \cite{eghbal2020} to gaming communities \cite{mnookin1996}. Governance documents in general are performative in the social scientific understanding of the term. Note that by ‘performative’, we do not mean that the constitutions are ornamental or do not serve a real purpose, rather that they enunciate the commitments of a group and serve to orient participants around a common project. For online communities, they also serve as a stable, anchoring feature of a space where the activities are fluid and participants can (typically) freely come and go.

\begin{figure}[htbp]
\centering
\includegraphics[width=0.5\linewidth]{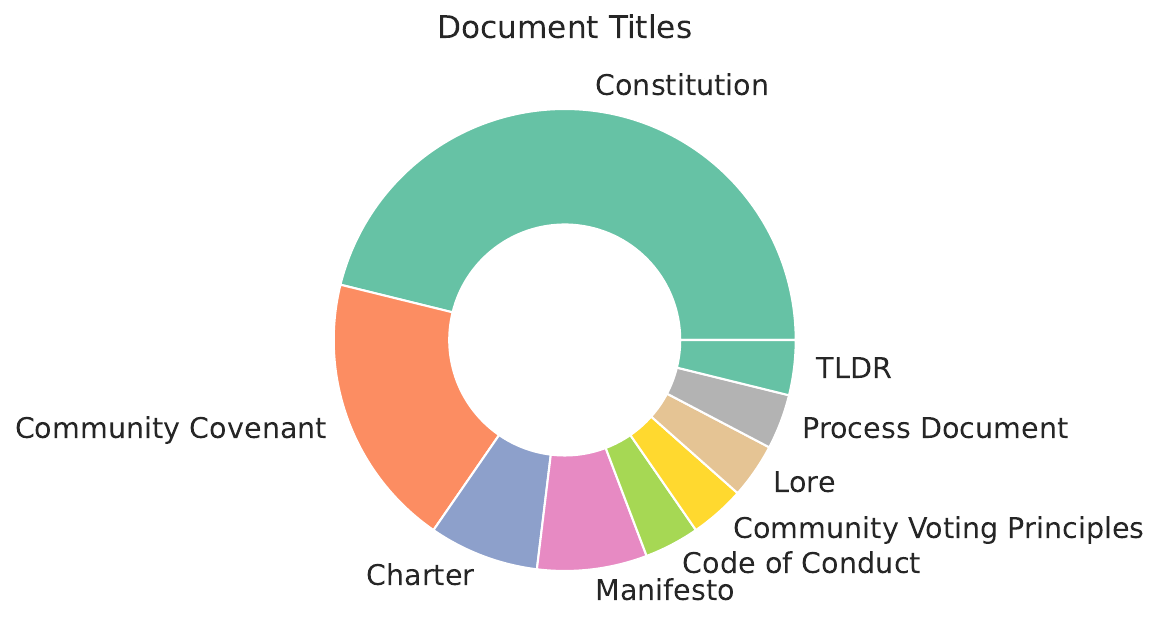}
\caption{Document titles in the analyzed data set.}
\label{fig:mylabel}
\end{figure}

We analyze governance documents as artifacts that use language to articulate ideas of technical and moral order specific to a project. A look at the language-specific elements of these constitutions can tell us about the social and moral dimensions of the projects they describe: how the community understands itself within an ecosystem of other actors, how it imagines its contribution, and against what forces it seeks to define itself.

The documents present a diversity of structures following a number of genres:

\begin{itemize}
    \item Some are modeled on the modern notion of the constitution with separate "articles" that define specific subsets of rights and responsibilities.

    \item Others emulate the manifesto form with an expression of values and intent along with a set of guidelines to follow.

    \item Some read more like documentation to a software project.
    \item Yet others appear to be narrating a hero's journey \cite{campbell2008}.
\end{itemize}

Each of these genres follow certain conventions and "do" certain things. Notably, they are more or less embedded in existing legal structures; for instance, the constitution of Ethereum World explicitly mentions Swiss law.

Beyond the structure or generic form of the documents themselves, the language of these constitutions further display a range of “registers,” or functional styles. Some are written in what would colloquially be called “legalese,” a register specialized to the legal profession which dumbfounds readers not versed in it. Others try to emulate “officialness” by using particles such as “shall” in an otherwise not specialized register. Some of the documents self-consciously marshal “kumbaya” type phrases and invocations, while others (notably, the ones following the 1Hive covenant template) are notably “plain,” that is, written in a register that is relatively unmarked by any specialized function. This has the effect of feeling the most inclusive, which perhaps explains why it has been emulated repeatedly.

These documents also invoke and address a number of different groups. The documents often speak in the voice of a first person plural, “we” (variously defined as “thinkers and doers,” “community of,” “members, contributors, leaders of”). They often articulate the intended beneficiaries of the project’s activities to be all of humanity (“all humans”, “humanity,” “individuals,” “the global society,” “everyone”), and they also frequently articulate the kinds of users and individuals which the project seeks to exclude. This is done implicitly by stating certain values (such as community over profit) or explicitly by naming (“malicious actors”). Future research may also wish to consider what the “we” represents within these documents, for example what is the intention of “we” and the function of this statement (for example, how the parameters of communities are made).

With some of the constitutions we observe references to other products and communities (in particular the ones that are based on the 1Hive community covenant) which locate the group at a particular place in a growing ecosystem. These references create intertextual links to other communities and governance practices, tools, and norms.

\subsection{Constitutions describe goals, values, and rights}

Following Redeker, et al. (2018) we identified the goals, values, and rights defined in each document (to the extent possible) and then coded them into overarching categories emergent from the texts. Our findings are significantly simpler compared to other frameworks for classifying digital rights and governance. This reflects both the relative simplicity of these early constitutional documents and the smaller size (and perhaps scope) of the political communities involved \cite{davies2014}.

\subsubsection{Goals}

\begin{figure}[htbp]
\centering
\includegraphics[width=0.5\linewidth]{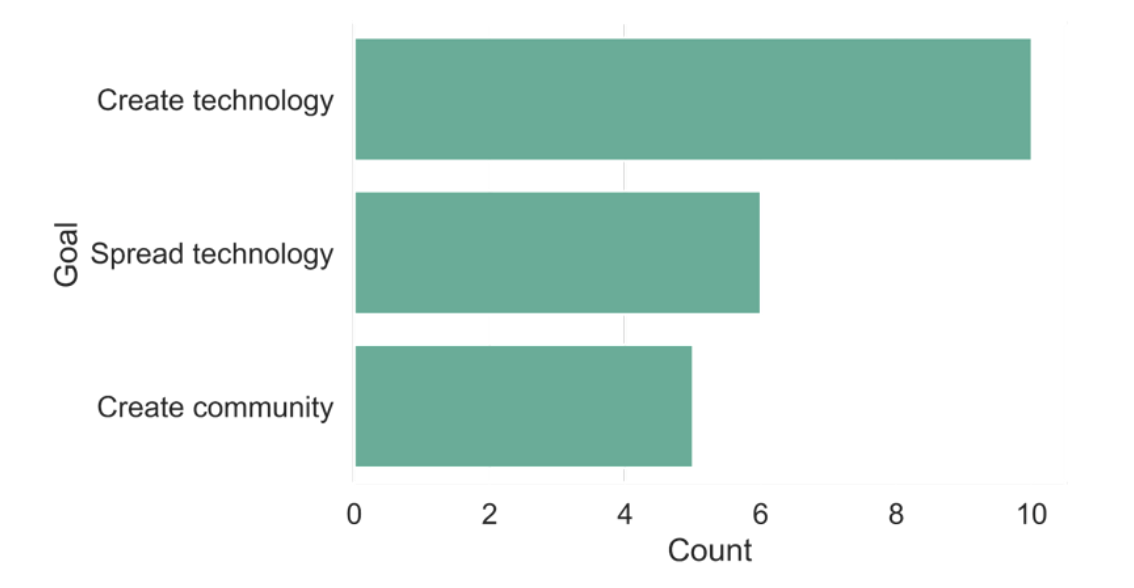}
\caption{Coded goals in the data set}
\label{fig:mylabel}
\end{figure}

\begin{itemize}
    \item \textbf{Create technology} - the community aims to develop specific technologies related to Web3.
    \item \textbf{Spread technology} - the community aims to spread in the sense of distribute this and related technologies in order to benefit the Web3 ecosystem.
    \item \textbf{Create community} - the community aims to build a community devoted to the other goals.
\end{itemize}

We have coded the goals that appear most frequently as “create technology,” “spread technology” and “create community.” These goals typify these communities as “recursive publics”: communities that come together in order to build the tools that allow them to come together as a community. In Kelty’s definition: “A recursive public is a public that is vitally concerned with the material and practical maintenance and modification of the technical, legal, practical, and conceptual means of its own existence as a public” \cite{kelty2008two}.

\subsection{Values}

\begin{figure}[htbp]
\centering
\includegraphics[width=0.5\linewidth]{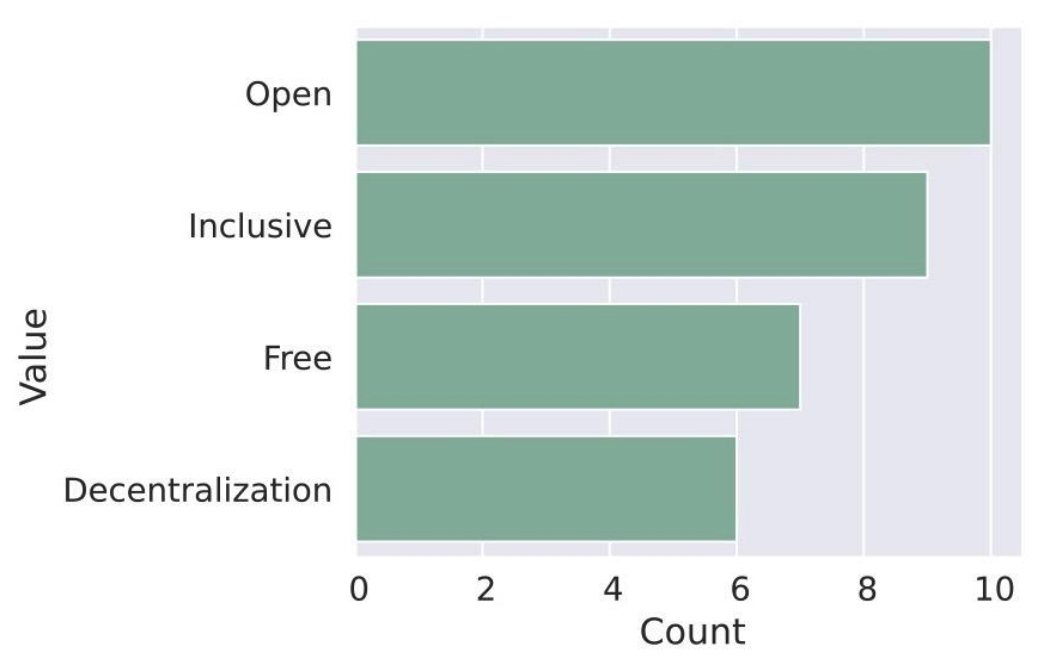}
\caption{Coded values in the data set}
\label{fig:mylabel}
\end{figure}

\begin{itemize}
    \item \textbf{Open} – allowing access to a product, community, environment.
    \item \textbf{Inclusive} – aiming to provide access to the community for people who might otherwise be excluded or marginalized.
    \item \textbf{Free} – without cost or payment; not under the control of another.
    \item \textbf{Decentralization} – not under centralized control, distributed.
\end{itemize}

The values we have listed above can be encountered in these documents, however there is no standard definition of what each of these values mean to each of these communities. Each of these values can be understood as an open signifier, meaning that different DAO communities will attach different meanings to the same term. The values “free” and “open” initially served as open signifiers in free and open-source software communities but have conceptually stabilized over time as participants debated their meanings  \cite{stallman2009}. We are not surprised to discover that, as inheritors of this tradition, Web3 communities would center “open” and “free” in their governance documents.

The value “inclusive” is notable because it echoes long-standing commitments to meritocratic inclusivity in open-source software communities, but does so in a sociopolitical context in which “inclusivity” refers to social justice/identity politics issues more specifically. For example, the 1Hive constitutions promise, quite prominently, the “right to a harassment-free experience”. Further research is needed to understand how “inclusivity” is modeled and practiced beyond the value stated in the constitution.


Decentralization, although not a novel term, has gained a new salience with blockchain’s features and affordances. One of the central premises of these documents is to make a claim to “decentralization”. Scholars have generally agreed that there is no common definition for decentralization \cite{schneider2019} and in the context of blockchain, “decentralization” is the “boundary object” \cite{star1989} that enables individuals from different communities to work together. What is interesting about decentralization is that it has become a newly emergent term that now encapsulates various goals, serving as a means toward an end. Commitments to decentralization in blockchain spheres have now evolved into sophisticated bodies of thoughts and ideas. DAO constitutions hence serve as artifacts that embed some of these ideas and commitments to decentralization. 

These documents provide a window into how communities understand decentralization and how its definition gets contested. The value of ‘decentralization’ hence dictates some of the conditions and precepts presented within these DAOs. It presents a particular narrative of how Web3 should operate (e.g. resistance to corporate power but not founder power). 

A cursory examination of the contextual uses of the values indicate that their meanings vary despite being presumably stable. “Open,” another floating signifier, does not mean the same thing in each document. Further analysis is needed to map out patterns in the semantic fields of these terms, the relationships of values to rights and goals and to the types of DAOs (in particular, whether the DAO has a financial goal or not).

\subsection{Rights}

\begin{figure}[htbp]
\centering
\includegraphics[width=0.8\linewidth]{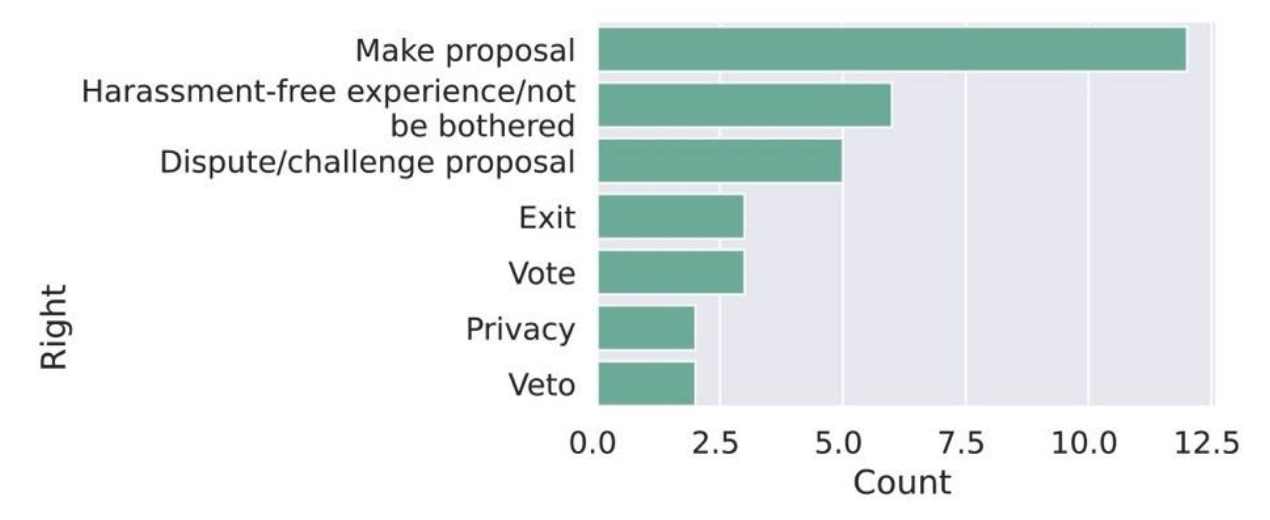}
\caption{Coded rights in the data set}
\label{fig:mylabel}
\end{figure}

\begin{itemize}
    \item \textbf{Right to create proposal} - The right to create proposals (usually by staking cryptocurrencies or governance tokens) 

    \item \textbf{Right to harassment free experience} - No person will be discriminated based on their age, body size, visible or invisible disability, ethnicity, sex characteristics, gender identity and expression, level of experience, education, socio-economic status, nationality, personal appearance, race, religion, or sexual identity and orientation.
    \item \textbf{Right to exit} - The right to exit a community as a participant pleases, typically including the ability to withdraw assets when leaving.
    \item \textbf{Right to vote} - The right to vote on proposals and other forms of referenda; sometimes to delegate that vote
    \item \textbf{Right to privacy} - The right to not have one’s personally-identifying information revealed, i.e. to be “doxxed”
    \item \textbf{Right to veto} - The right to veto a proposal
\end{itemize}

Rights directly shape behavior by specifying permissible actions. In this sense, more so than values or goals, rights are highly consequential for governance. Within DAOs, rights are further distinct because of their correspondence with smart contracts. Because they do not possess the expressive power of language, smart contracts are a poor medium for capturing values or goals. However, since they can more directly alter the state of the world, smart contracts are highly promising as a means of enforcing rights.

Within the documents in our dataset, ‘rights’ are defined more uniformly and precisely than ‘values’ or ‘goals’. The right to create proposals, for instance, refers to an action that is much more specific than, for example, than ‘decentralization’ or ‘open’ as a stated value. We hypothesize that this has to do with the affordances of the smart contract; in order to interact with a smart contract, a right needs to be specified in technically-precise language. In other words, rights have to be translated into code to be put on-chain, and vice versa, a right that already exists on-chain can be translated into precise language. A more prosaic reason for the uniformity of stated rights is that there is substantial copying and forking between DAO constitutions, e.g. a few constitutions in our data set were derived from 1Hive’s constitutional template.

As the social and technical infrastructure around digital governance improves, we hope and expect to see the breadth of rights broaden. The rights so far are few in number and similar in theme. We hope to see rights become broader and more useful.

\section{Part III: Writing DAO constitutions}

In this section, we make a set of practical recommendations on how to draft DAO constitutions in order to maximize their use in governance. In particular, we recommend that constitutions be digital, amendable, short, expository, and early.

\begin{itemize}
  \item \textbf{Digital.} The constitution should be accessible at a URI, stored in an accessible digital format such as a \texttt{.md} or \texttt{.txt} file, and make use of hyperlinks, especially links to any example policies, smart contracts, digital platforms, or other DAOs relevant to the community.
  \item \textbf{Amendable.} There is an accessible, transparent process for amending or changing the constitution. Often, but not always, this means having a section dedicated to amendment procedures.
  \item \textbf{Short.} The constitution should be short and focused. Do not overspecify definitions, rules, and processes. As a rule of thumb, imagine a document that more than 50\% of your community would actually read. That might be three tweets, or it might be a page. It is not a 30-page legal document.
  \item \textbf{Expository.} Each goal, value, or right should come with enough context and exposition, often through an example, so that any member of the community can grasp the concept.
  \item \textbf{Early.} Constitutions should be written and promulgated at the beginning of an organization or community’s creation.
\end{itemize}

These recommendations are based on a series of interviews with the drafters of DAO constitutions, including those from DAOhaus, the Ethereum Name Service (ENS), and the Token Engineering Commons (TEC), and on the empirical practices analyzed in Part II. Along with these recommendations, we provide a template constitution along with a forkable code repository to help DAOs draft and maintain such constitutions (see the Appendix).

\section{Part IV: Towards computational constitutionalism}

A digital constitution, extending Redeker et al.’s definition, is a text that articulates a set of political rights, governance norms, and limitations on the exercise of power within an online community. So far, we have created a data set of digital constitutions (among other governance documents) of DAOs and related Web3 entities, coded the goals, values, and rights present in these texts, identified their major functions and themes, and made several recommendations for useful constitutions based on our findings. But as we noted in the introduction, DAOs are defined by their usage of smart contracts. These smart contracts are computational constitutions insofar as they encode rights, norms, and limitations on the exercise of power within the DAO \cite{zargham2021}. We refer to the broader pattern of using computational artifacts to encode constitutional rights and processes as \textit{computational constitutionalism}.

\subsection{Relating digital and computational constitutions}
While there are clear patterns across digital constitutions, those texts are still relatively bespoke to the communities that write them. In contrast, the smart contracts used by most DAOs are not bespoke; the specialized technical knowledge required to write functional and secure smart contracts precludes most DAOs from creating theirs from scratch. As a result, a number of templates and "DAO factories", produced by external organizations, have emerged to help DAO creators deploy the core smart contracts of a DAO with a specified set of configurable parameters. The distribution of templates used by DAOs in our dataset is shown below.

\begin{figure}[htbp]
\centering
\includegraphics[width=0.8\linewidth]{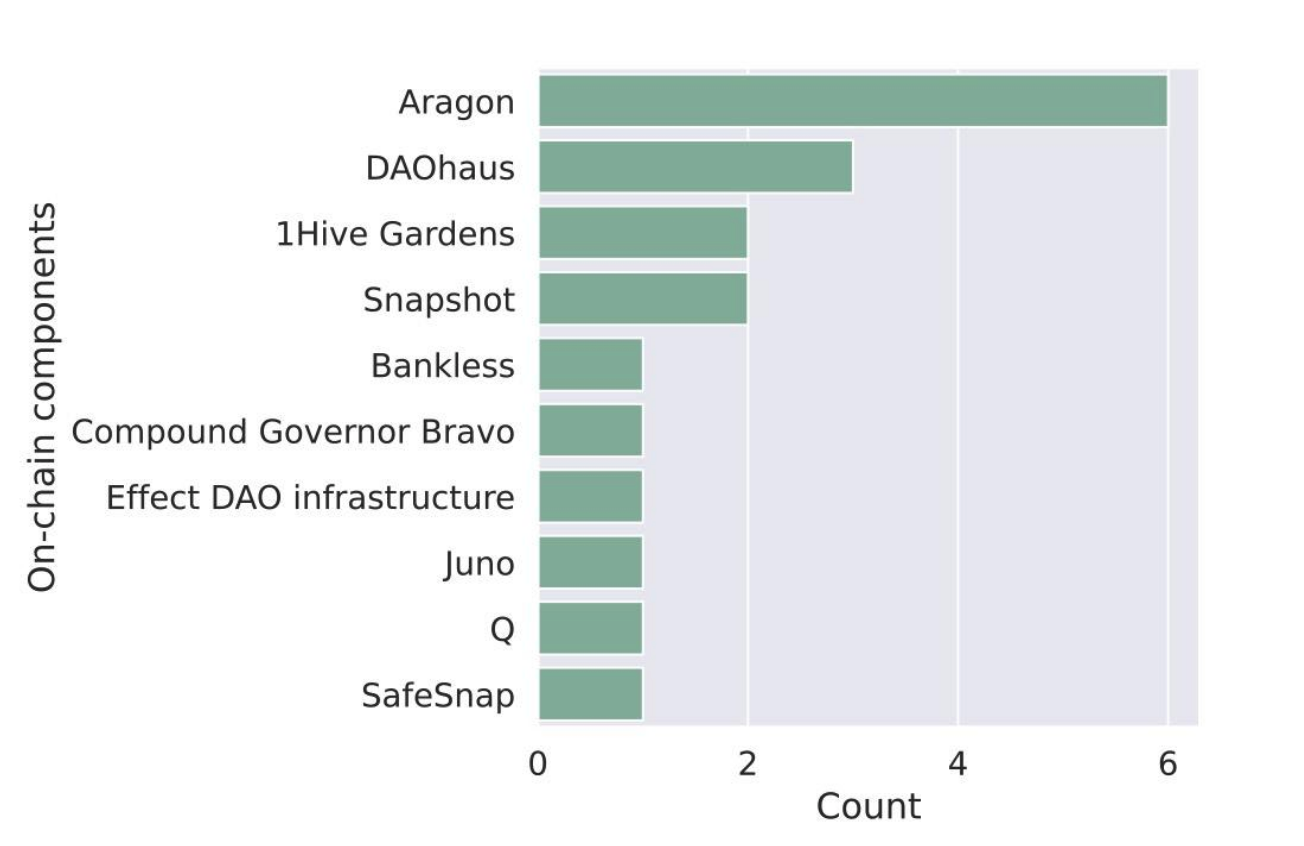}
\caption{Count of on-chain frameworks used by DAOs in the data set. Note that this is incomplete as some constitutions correspond to blockchains whose governance features are embedded in the consensus protocol rather than in a smart contract.}
\label{fig:mylabel}
\end{figure}

Much work remains to be done to understand the relationship between these smart contracts and “off-chain” institutional constructs such as a constitution. While smart contracts neither need nor assume constitutions (and vice versa), the value of understanding DAOs through the lens of computational constitutionalism is to see how certain rights and certain forms of governance can be more easily and effectively provided through a smart contract, by a constitution (and constitutional norms), or by some combination of the two. As we have seen, the textual constitutions commonly articulate rights already guaranteed by the smart contracts, including proposal creation and proposal dispute as well as voting, veto, and exit. But to what extent should constitutions reference the smart contract? In what cases should a written constitution or charter defer to the contract, and in what cases should the contract defer to the constitution? 

To give some early insight into these questions, we report on an early case study of computational constitutionalism within the Cardano ecosystem.
 
\subsection{An early case study}
Cardano is a layer one blockchain comparable to Bitcoin or Ethereum. Released in 2017, it is the fourth largest blockchain by market capitalization (\$16B as of Sep. 1, 2023). In early 2023, as part of their planned “Voltaire” upgrade, Cardano sought to decentralize its governance by allocating more resources and responsibilities away from Input Output Global Inc. (IOG), a private company charged with the development and maintenance of Cardano, to a constitutional system controlled by different stakeholders within the Cardano ecosystem. In particular, IOG sought to do this through two documents: a technical specification of a software upgrade to the Cardano protocol, CIP-1694 \cite{corduan2023voltaire}, as well as a formal, community-facing constitution.

Over the course of several months starting in April 2023, we observed and helped advise aspects of Cardano’s constitution drafting process. To add additional context, we also interviewed both employees of IOG as well as other participants in the process. As of the writing of this article, Cardano’s constitutional drafting process is still underway, and the full draft constitution is not yet public. We base this report on CIP-1694 as well as elements of the constitution, including principles for its drafting (see Appendix B), that have been presented publicly.

First, it was interesting to observe and understand how the Cardano team interpreted the recommendations we made above; their stated principles for the constitution were “digital, participatory, short, explanatory, and alignment” (see Appendix B for the full definitions of what these mean), compared to “digital, amendable, short, expository, and early”. In effect, the largest changes (removal of “amendable”, addition of “participatory” and “alignment”) all concern the long-term sustainability of the ecosystem; the amendable condition poses sustainability as a function of the ecosystem’s ability to adapt, while Cardano’s alignment condition poses sustainability as a function of the ecosystem’s ability to align user motivations, especially economic ones, with ecosystem values (and vice versa): “The Cardano constitution must support a minimum viable governance system that both encourages and facilitates long-term, active participation but equally recognizes the economic motivations that are likely to contribute to participation. It must also provide the roadmap for aligning the community and ecosystem’s core values.”

\emph{To what extent should constitutions reference the smart contract?} CIP-1694 specifies a set of roles with rights enforced on-chain, including a Constitutional Committee, “DReps” (for delegated representatives), and a set of stake pool operators (SPOs). CIP-1694 also references other features of the Cardano protocol developed in other technical documents, in particular the ADA token and ADA token holders. However, it explicitly describes as out of scope anything to do with the Cardano constitution. 
Current evidence suggests that the constitution will be drafted with CIP-1694 in mind, with a relative focus on articulating values than the architecture of roles, which is already specified in CIP-1694.



\emph{In what cases should a written constitution or charter defer to the contract, and in what cases should the contract defer to the constitution?} 
As specified in CIP-1694, the Cardano constitution serves as a set of guidelines for the decisions of the Constitutional Committee, which will have the power to shape decisions on changes to the Cardano protocol (though only in part, as changes to the client code need to be officially ratified by the stake pool operators that operate the actual infrastructure). Ergo, much of the role of the constitutional text is already specified by the technical architecture of Cardano. However, CIP-1694 also specifically notes places where the constitution takes precedence, for example in the management of off-chain legal and financial entities. 
While the protocol’s operation may in principle be inured to these entities, depending on the degree of decentralization, in practice the off-chain factors can sharply constrain the incentives of different operators on the blockchain.

\section{Future work}
Computational constitutionalism is important because it suggests a very general pattern—code combined with text—for how governance will evolve within digital organizations. By understanding the governance of today’s DAOs, we can better understand how to design future organizations and communities, both online and offline. While early case studies can give us useful intuitions for how smart contracts interact with constitutions, more study is needed to ground more robust and rigorous recommendations. To understand how smart contracts function on their own and in conjunction with textual constitutions or other governance documents and processes, we must investigate three distinct data sets:
\begin{enumerate}
    \item the rights and affordances defined within DAOs' smart contracts, 
\item the parameter configurations with which DAOs have chosen to deploy their own computational governance systems from template smart contracts, and 
\item the actual activity and outcomes of governance circumscribed by constitutions and/or smart contracts. 

\end{enumerate}

These data sets will allow us to investigate what framework designers have considered necessary and sufficient to implement governance on a blockchain, what DAOs leaders have considered necessary and desirable to implement governance on a blockchain, and what governance actions DAO members actually take within the constraints of the governance configuration. Ultimately, understanding these will allow for the creation of theoretical frameworks and practical tools for the design and governance of online communities. 

Beyond analyzing smart contracts, there is still much work to be done to expand the dataset of textual documents and to contextualize their creation and use. We hope to continue building our dataset not only by lowering the friction of contributing data but also by automating certain aspects of data collection, especially through improvements to our template or to the associated tools and widgets in the code repository.

Additionally, further information is needed to understand the relationship between the structure and contents of a textual constitution and its observed usage in governing a community. For what audiences was the constitution written, and how does its publishing location relate to these? What are the processes by which the constitution was created, and who was involved in its drafting and revision? In what ways and with what frequency is the constitution being referenced and revisited by the community it governs?

\section{Acknowledgements}
We would like to thank Jacky Zhao, Jasmine Wang, and Evan Miyazono. This work was made possible through a grant from the Filecoin Foundation. Joshua Tan was also funded through the EPSRC IAA Doctoral Impact Scheme and the Stanford Digital Civil Society Lab.

\bibliographystyle{unsrt}
\bibliography{web3_constitutions}

\appendix
\section{Appendix A: Template and repository}

We have packaged the template below, along with a few helpful tools, analytics, and the current raw data set of DAO constitutions, into a code repository hosted on GitHub, a well-known code-sharing platform. This repository can be found at \url{https://github.com/metagov/constitution-template}.
The coded data set can be found at \url{https://airtable.com/shrvSk01p3E1wwH77}.
We have published this template, along with a version of this article, at a public website (currently hosted at \url{https://constitutions.metagov.org}).
The code for that website can be found at \url{https://github.com/verses-xyz/constitutions}.

\begin{lstlisting}
# Constitution of <Organization>

Constitutions following this template are _digital_, _amendable_, _short_, _expository_, and _early_. 

- _Digital._ The constitution should be accessible at a URI, stored in an accessible digital format such as a .md or .txt file, and make use of hyperlinks, especially links to any example policies, smart contracts, digital platforms, or other DAOs relevant to the community. 
- _Amendable._ There is an accessible, transparent process for amending or changing the constitution. Often, but not always, this means having a section dedicated to amendment procedures. 
- _Short._ The constitution should be short and focused. Do not overspecify definitions, rules, and processes. As a rule of thumb, imagine a document that more than 50\% of your community would actually read. That might be three tweets, or it might be a page. It is not a 30-page legal document. 
- _Expository._ Each goal, value, or right should come with enough context and exposition, often through an example, so that any member of the community can grasp the concept. 
- _Early._ Constitutions should be written and promulgated at the beginning of an organization or community's creation.

## Preamble
<!-- The preamble introduces the community / DAO, its goals, and its values. Focus on just the 2-3 values and 2-3 goals that really matter. -->

## Article 1
<!-- Each article of the constitution should respond to the goals and values articulated in the preamble. Each article should address an important issue, policy, institution, or right. -->

## Article 2:
## Article 3:
## Article 4:
## Article 5:
<!-- We recommend that new constitutions begin with five or fewer articles. Additional articles can be added through amendments. This is also a good opportunity to practice going through the amendment process! --> 

<!-- We encourage communities to fill in the following metadata as a a comment directly in the constitution's .md file. --> 

<!-- \{ "@context": "https://constitutions.metagov.org", 
"type": "constitution", 
"title": "<title of the document>",
"communityName": "<name of the DAO>",
"daoURI": "<URI of daoURI, see DAOIP-2>",
"author": "<names of authors, comma-separated>"
"discussions-to": "<URI>"
"created": "<YYYY-MM-DD>",
"lastModified": "<YYYY-MM-DD>",
"previousConstitutionURI": "<URI>",
"<URI>", "inForce": "<True, False>" \} 
--> 

<!-- Metagov Metaconstitutions v1.1. Released under a CC0 License. -->
\end{lstlisting}

\section{Appendix B: Cardano Constitution}

The following are the principles of the Cardano constitution as articulated during the constitutional drafting process:






\textbf{Initial Thoughts on Constitution Principles}

Digital. 
The constitution should be accessible in a digital format.  Ideally, it should be contained in or linked to the blockchain itself. In the future it might be converted to code.

Participatory.
To encourage community involvement, processes for community involvement through voting, delegation of authority to third-party experts and the process of future amendments to the constitution itself must all work in connection with on-chain voting procedures.  In particular, the constitution should serve as the foundational platform for governance in a decentralized ecosystem.

Short. 
Participation in a blockchain ecosystem is voluntary and may be transitory.  The institutions supporting a blockchain ecosystem will not likely be static.  Therefore, we believe the constitution should be targeted and focused on overarching goals of ease of democratic participation and experimentation, as well as memorializing the core philosophy underlying the Cardano protocol.

Explanatory. 
Each goal, value, or right should come with enough context and exposition, to facilitate easy understanding and encourage the involvement and buy-in of a diverse universe of participants in the Cardano ecosystem.  

Alignment. 
As discussed above, citizens of a country are “stuck” with their constitution unless they can overthrow their government or can flee to another jurisdiction.  There is no such stickiness for a blockchain ecosystem.  The Cardano constitution must support a minimum viable governance system that both encourages and facilitates long-term, active participation but equally recognizes the economic motivations that are likely to contribute to participation. It must also provide the roadmap for aligning the community and ecosystem’s core values.

\end{document}